# The effect of anisotropy and external magnetic filed on the thermal entanglement in a two-spin-qutrit system

S.Y. Mirafzali , M. Sarbishaei,
Department of Physics, Ferdowsi University of Mashhad, Iran

We study the thermal entanglement in a two-spin-qutrit system with anisotropy in the exchange coupling between two spins. We use the realignment criterion to distinguish the entangled states, and the negativity for measuring the entanglement in this system. We see that the anisotropy can provide an additional parameter for enhancing the entanglement.

## 1. Introduction

Entanglement is a non-local correlation between quantum systems that does not exist classically. It plays a central role in quantum information processing such as quantum teleportation, dense coding etc.

The entanglement properties of various systems have been intensively studied, but spin 1/2 systems have been considered in most of these studies and higher spin systems have been less investigated. The reason for this restriction is that a good operational entanglement measure, when the system is in a mixed state, has been known only for two spin-halves systems. However there is currently an effort to study entanglement in systems with higher spins.

There are several preceding works on entanglement in spin-one chains [1-7]. Zhang et al [8] have been studied the thermal entanglement of a system consists of two wells in the optical lattice with one spin-1 atom in each well. One can summarize the Hamiltonian of this system as follow [9].

$$H = \varepsilon + J(S_1 \cdot S_2) + K(S_1 \cdot S_2)^2 \quad (1)$$

The above Hamiltonian, in the absence and in the presence of an external magnetic field B, has been studied in detail in [10-13]. However, the effect of anisotropy in exchange coupling between two atoms has not been considered. It can provide an additional parameter (besides the other parameters) for controlling entanglement. In this paper we study the effects of anisotropy on thermal entanglement.

## 2. Negativity and realignment criterion.

As it was noticed by Peres [14], a necessary criterion for mixed state of a bipartite system to be separable is that its partial transpose with respect to one of the subsystems is positive. Horodecki family [15] have shown that this condition is also sufficient if the Hilbert space of the bipartite system has dimension $2 \times 2$ or $2 \times 3$. But for larger dimensions, the PPT (or Peres-Horodecki) criterion is not a sufficient one. Nevertheless, the negative partial transpose (NPT) gives a sufficient condition for entanglement.

Vidal and Werner [16] introduced a computable measure of entanglement called negativity. The negativity of a state $\rho$ is defined as

$$N(\rho) = \sum_i |\mu_i|$$

where $\mu_i$ is the negative eigenvalue of $\rho^{T_1}$, and $T_1$ denotes the partial transpose with respect to the first system. Negativity essentially measures the degree to which $\rho^{T_1}$ fails to be positive. Therefore, it can be seen as a quantification of the PPT criterion for separability and because of that we cannot confirm whether the state is entangled or not in the region of zero negativity. We must seek for another criterion that reveals entanglement where the negativity fails to distinguish between separable states and entangled states.

Another computable separability criterion was introduced by Rudolph [17] and Chen et al [18] called realignment criterion. It is independent of Peres criterion, and turns out to be strong enough to detect entanglement of almost all known states, for which the former criterion fails [19]. However, for some states it is weaker than PPT one [18].

In order to make use of this criterion we only need to rearrange the entries of $\rho$. Then the sum of the square roots of the eigenvalues for $\tilde{\rho}\tilde{\rho}^\dagger$ should be $\leq 1$, or equivalently

$$R = \log\left(\sum_i \sigma_i\right) \leq 0$$

where $\sum_i \sigma_i$ is the sum of the square roots of the eigenvalues for $\tilde{\rho}\tilde{\rho}^\dagger$. The realigned form of a $m \times m$ block matrix $\rho$ with block size $n \times n$ is defined as

$$\tilde{\rho} = \begin{bmatrix} vec(\rho_{1,1})^T \\ \vdots \\ vec(\rho_{m,1})^T \\ \vdots \\ vec(\rho_{1,m})^T \\ \vdots \\ vec(\rho_{m,m})^T \end{bmatrix}$$

where for each $m \times n$ matrix $A = [a_{ij}]$ ($a_{ij}$ is the matrix entry of $A$), $vec(A)$ is defined as

$$vec(A) = [a_{11},...,a_{m1},...,a_{12},...,a_{1n},...,a_{mn}]^T$$

### 3. The model's Hamiltonian and thermal entanglement

We consider the Hamiltonian of the form

$$H = J(S_{1x}S_{2x} + S_{1y}S_{2y} + \Delta S_{1z}S_{2z}) + K(S_{1x}S_{2x} + S_{1y}S_{2y} + \Delta S_{1z}S_{2z})^2 + B(S_{1z} + S_{2z})$$

where $S_\alpha (\alpha = x, y, z)$ are the spin operators, $J$ is the strength of Heisenberg interaction, $K$ is the nonlinear couplings constant and the magnetic field is assumed to be along the z-axes. We also have considered an anisotropy in the z direction that is represented by $\Delta$. When the spin for each site $S_i = 1$ ($i = 1,2$), its component take the form

$$S_{ix} = \tfrac{1}{\sqrt{2}}\begin{pmatrix} 0 & 1 & 0 \\ 1 & 0 & 1 \\ 0 & 1 & 0 \end{pmatrix}, \quad S_{iy} = \tfrac{1}{\sqrt{2}}\begin{pmatrix} 0 & -i & 0 \\ i & 0 & -i \\ 0 & i & 0 \end{pmatrix},$$

$$S_{iz} = \tfrac{1}{\sqrt{2}}\begin{pmatrix} 1 & 0 & 0 \\ 0 & 0 & 0 \\ 0 & 0 & -1 \end{pmatrix}$$

The state of the system described by the Hamiltonian $H$ at thermal equilibrium is:

$$\rho(T) = \exp(-\beta H)/Z,$$

where $Z = Tr[\exp(-\beta H)]$ is the partition function and $\beta = 1/K_B T$ ($K_B$ is the Boltzmann's constant that for simplicity being set to be unit, $K_B = 1$, and $T$ is the temperature). As $\rho(T)$ represent a thermal state, the entanglement in the state is called the thermal entanglement. We analyze the dependence of thermal entanglement in this system on temperature and external field and study the effects of anisotropy on the thermal entanglement.

### 4. The solutions

We considered two cases:
**Case 1:** $K = 0$
The eigenvalues and eigenvectors of $H$ are obtained as

$H|1,1\rangle = (J\Delta + 2B)|1,1\rangle,$

$H|-1,-1\rangle = (J\Delta - 2B)|-1,-1\rangle,$

$H|\Psi_1^\pm\rangle = (B \pm J)|\Psi_1^\pm\rangle,$

$|\Psi_1^\pm\rangle = \tfrac{1}{\sqrt{2}}(|1,0\rangle \pm |0,1\rangle),$

$H|\Psi_2^\pm\rangle = (-B \pm J)|\Psi_2^\pm\rangle,$

$|\Psi_2^\pm\rangle = \tfrac{1}{\sqrt{2}}(|0,-1\rangle \pm |-1,0\rangle),$

$H|\Phi^\pm\rangle = \xi^\pm|\Phi^\pm\rangle,$

$|\Phi^\pm\rangle = [8 + (\eta^\pm)^2]^{-1/2}(2|1,-1\rangle + \eta^\pm|0,0\rangle + 2|-1,1\rangle),$

$\xi^\pm = \tfrac{1}{2}(-J\Delta \pm J\sqrt{\Delta^2 + 8}), \quad \eta^\pm = \Delta \pm \sqrt{\Delta^2 + 8},$

$H|\Phi\rangle = -J\Delta|\Phi\rangle,$

$|\Phi\rangle = \tfrac{1}{\sqrt{2}}(|1,-1\rangle - |-1,1\rangle),$

where $|i,j\rangle$, $(i,j = 1, 0, -1)$ are the eigenstates of $S_{1Z}S_{2Z}$.

We calculated numerically the negativity and $R = \log\left(\sum_i \sigma_i\right)$ of $\rho(T)$. For the case of $J = -1$, we plotted negativity and $R$ as a function of $B$ and $\Delta$ for four typical temperatures and the results are shown in Fig 1.

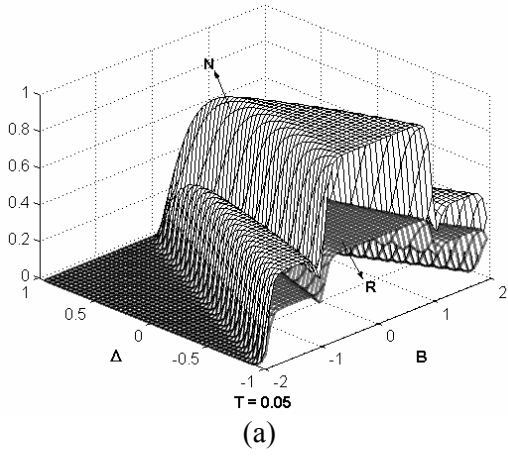

(a)

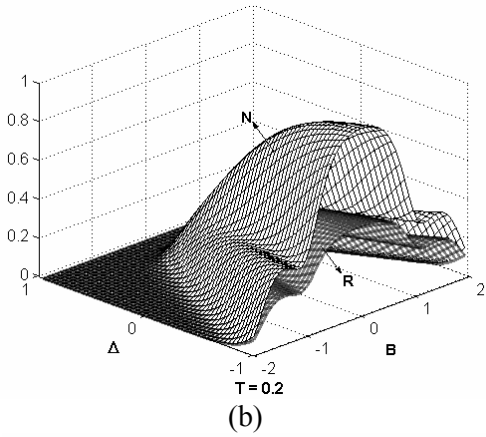

(b)

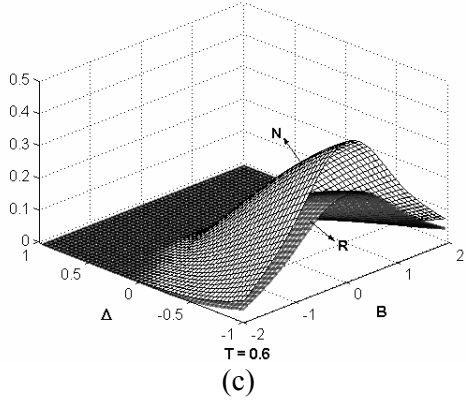

(c)

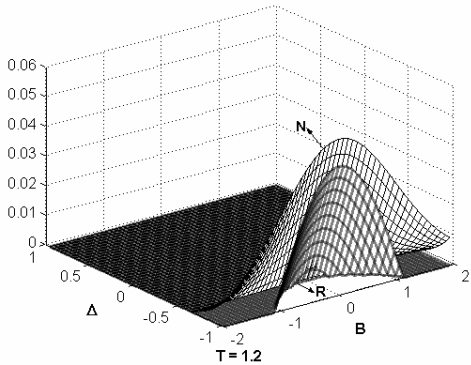

Fig.1. The negativity ($N$) and $R$ as a function of $B$ and $\Delta$ at (a) $T = 0.05$, (b) $T = 0.2$, (c) $T = 0.6$, (d) $T = 1.2$.

We see that almost $N$ (negativity) and $R$ behave alike. They are even functions of $B$ and their maximum is at $B = 0$ (for fixed $\Delta$ and $T$). At $T = 0.05$ (fig.1 (a)), they have three sharp peaks. The center of the middle peak locates at $B = 0$, but with increasing the temperature the left and the right peak disappear (fig.1 (b)-(d)). In general, as the magnetic field and the temperature increase, $N$ and $R$ decrease. At low temperatures, $N$ and $R$ are as strong as each other, but at higher temperatures ($T > 0.2$), $N$ is stronger than $R$ (fig.1 (d)). The threshold temperature, at which the negativity and $R$ vanish, is $T \approx 1.36667$.

The maximum values of $N$ and $R$ (for fixed $B$ and $T$) occur when $\Delta$ is equal to $J$ (see fig 2), and they vanish as the anisotropy ($\Delta$) tends to $-J$ (in figure.2, $-J = 1$). For $\Delta = J$, in the interval $-3 \leq B \leq 3$, $N$ and $R$ are nonzero, but as $\Delta$ tends to $-J$, this interval become smaller. For $J = 1$, we obtain the same results as $J = -1$

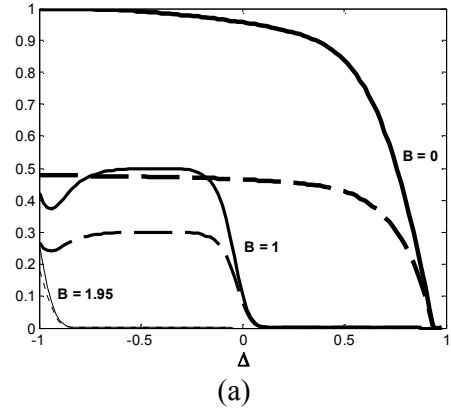

(a)

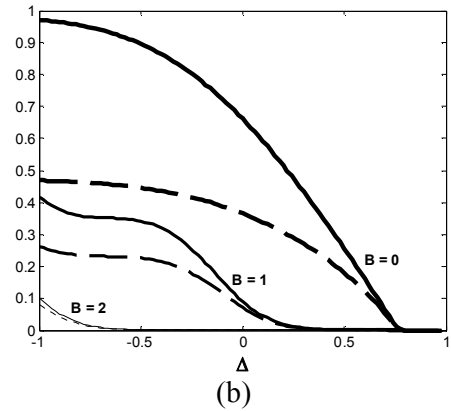

(b)

Fig.2. The negativity (solid lines) and $R$ (dashed lines) as a function of $\Delta$ for three typical $B$ at (a) $T = 0.05$, (b) $T = 0.2$

**Case 2:** $K \neq 0$

The eigenvalues and eigenvectors of $H$ are obtained as

$H|1,1\rangle = (J\Delta + 2B + K\Delta^2)|1,1\rangle$,

$H|-1,-1\rangle = (J\Delta - 2B + K\Delta^2)|-1,-1\rangle$

$H|\Psi_1^\pm\rangle = (B \pm J + K)|\Psi_1^\pm\rangle$,

$|\Psi_1^\pm\rangle = \frac{1}{\sqrt{2}}(|1,0\rangle \pm |0,1\rangle)$

$H|\Psi_2^\pm\rangle = (-B \pm J + K)|\Psi_2^\pm\rangle$,

$|\Psi_2^\pm\rangle = \frac{1}{\sqrt{2}}(|0,-1\rangle \pm |-1,0\rangle)$

$H|\Theta^\pm\rangle = \frac{\zeta^\pm}{2}|\Theta^\pm\rangle$,

$|\Phi^\pm\rangle = \frac{\left(2|1,-1\rangle - \frac{\zeta^\mp}{J - K\Delta}|0,0\rangle + 2|-1,1\rangle\right)}{\left[8 + (\zeta^\mp)^2\right]^{\frac{1}{2}}}$

$\alpha = -J\Delta + K\Delta^2 + K$,

$\zeta^\pm = \alpha + K \pm \sqrt{(\alpha + K)^2 + 8(J - K\Delta)^2}$

$H|\Phi\rangle = (-J\Delta + K\Delta^2)|\Phi\rangle$,

$|\Phi\rangle = \frac{1}{\sqrt{2}}(|1,-1\rangle - |-1,1\rangle)$

We calculated numerically the negativity and $R$ of $\rho(T)$. In the former case, we saw the effect of temperature on entanglement. As we increase the temperature, $N$ and $R$ decrease. In this case the effect of temperature is the same as former one, therefore the temperature is taken constant in the following discussion ($T = 0.2$).

For the case of $J = -1$, we plotted the negativity and $R$ as a function of $K$ and $\Delta$ for four typical $B$ and the results are shown in Fig 3 ($N$ and $R$ are even functions of $B$ therefore we only considered $B \geq 0$).

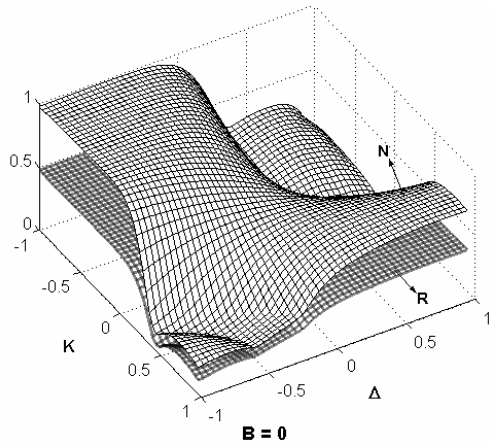

B = 0

(a)

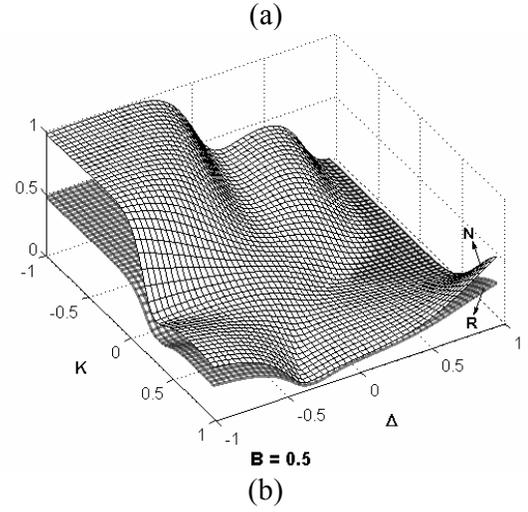

B = 0.5

(b)

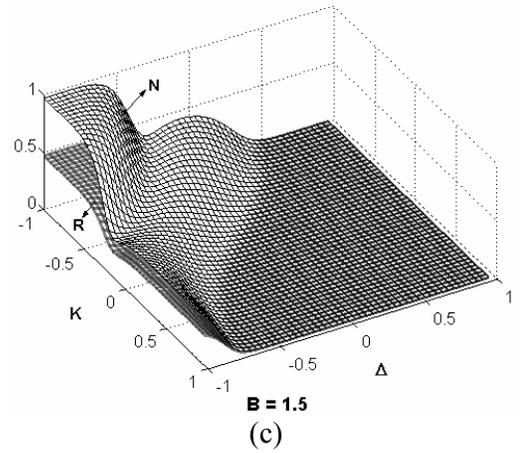

B = 1.5

(c)

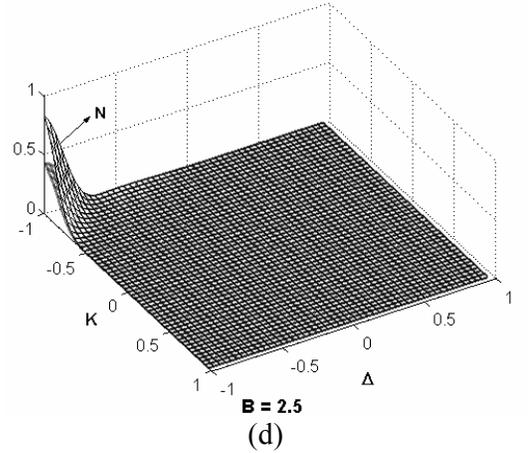

B = 2.5

(d)

Fig.3. The negativity ($N$) and $R$ as a function of $K$ and $\Delta$ at (a) $B = 0$, (b) $B = 0.5$, (c) $B = 1.5$, (d) $B = 2.5$ ($T = 0.2$).

We consider four regions:
**1.** The sign of $J$, $K$ and $\Delta$ are the same: In this region, for fixed $B$, the values of $N$ and $R$ are larger than their values in the other regions (their maximum is at $J = K = \Delta$). They monotonously decrease as $B$ increases.

**2.** The sign of $K$ and $J$ are the same but not the same as $\Delta$: as $B$ increases from zero to $B = 0.5$, $N$ and $R$ increase, then they decrease as $B$ increases further.

**3.** The sign of $K$ and $\Delta$ are the same but not the same as $J$: in this case, $N$ and $R$ monotonously decrease as $B$ increases. They vanish for $B > 0.8$. The maximum value of $N$ and $R$ (which occurs at $B = 0$) is about 0.7 and 0.4 respectively.

**4.** The sign of $\Delta$ and $J$ are the same but not the same as $K$: This region is similar to region 2. But, in this region, at some points, $N$ is stronger than $R$.

For the case of $J = 1$, we obtained the same results as $J = -1$.

## Conclusions

We used the negativity, to measure entanglement in this system. We also used the realignment criterion to distinguish between separable states and entangled states. However, we see that in this system, the realignment criterion is weaker than negativity.

We saw when $J$, $K$ and $\Delta$ are equal, we have maximum entanglement. In general, as the external magnetic field and the temperature increase, negativity and $R$ decrease.

## Acknowledgments

We would like to tank R.A.Zeinali for his valuable experience with MATLAB.